\documentclass[preprint,12pt,number]{elsarticle}

\usepackage{amssymb}
\usepackage{amsmath}

\usepackage{dsfont}
\usepackage{xcolor}
\usepackage{graphicx}
\usepackage{bm}
\usepackage[margin=1in]{geometry}
\usepackage[colorlinks=true,citecolor=magenta,linkcolor=blue]{hyperref}
\newcommand{\erre}{\mathbb{R}}
\newcommand{\be}{\begin{equation}}
\newcommand{\ee}{\end{equation}}
\newenvironment{equations}{\equation\aligned}{\endaligned\endequation}
\newcommand{\fer}[1]{(\ref{#1})}
\newcommand{\x}{\mathbf{x}}

\newcommand{\eq}{\mathrm{eq}}

\newcommand{\R}{\mathbb{R}}

\newcommand{\dd}{\mathrm{d}}

\journal{Physica A}

\begin{document}

\begin{frontmatter}



\title{Measuring inequality in society-oriented\\ Lotka--Volterra-type kinetic equations  } 


\author[aff1]{M. Menale} 
\author[aff2]{G. Toscani} 

\affiliation[aff1]{organization={Department of Mathematics and Applications, University of Naples Federico II, Italy},
            }
\affiliation[aff2]{organization={Department of Mathematics, University of Pavia, and Institute of Applied Mathematics and Information Technology ,  Pavia, Italy}
}

\begin{abstract}
We present a possible approach to measuring inequality in a system of coupled Fokker–Planck-type equations that describe the evolution of distribution densities for two populations interacting pairwise due to social and/or economic factors. The macroscopic dynamics of their mean values follow a Lotka–Volterra system of ordinary differential equations.
Unlike classical models of wealth and opinion formation, which tend to converge toward a steady-state profile, the oscillatory behavior of these densities only leads to the formation of local equilibria within the Fokker–Planck system. This makes tracking the evolution of most inequality measures challenging. However, an insightful perspective on the problem is obtained by using the coefficient of variation, a simple inequality measure closely linked to the Gini index. Numerical experiments confirm that, despite the system's oscillatory nature, inequality initially tends to decrease.
\end{abstract}

\begin{highlights}
\item We measure inequality in a system of coupled Fokker--Planck-type equations describing the evolution of the distribution densities of a multi-agent society interacting with the bank system via deposits and loans.
\item  The evolution of the mean values is shown to obey to a Lotka--Volterra system.

\item A comparison between the oscillations of the mean values and those of the coefficients of variation shows that the evolution of the system leads to a reduction of the inequality of the multi-agent society.
\end{highlights}

\begin{keyword} Lotka--Volterra equations \sep Fokker--Planck system \sep Quasi--equilibria \sep Inequality measures \sep Coefficient of variation.


\begin{MSC}
35Q20 \sep 35Q84 \sep 91B80
\end{MSC}

\end{keyword}

\end{frontmatter}

\section{Introduction}
The description of social phenomena in a multi-agent system can be successfully achieved by resorting to methods of statistical physics, and,  in particular, to kinetic theory. In presence of an infinite number of agents, that justify the kinetic approach, the main goal of mathematical modelling is to build master equations of Boltzmann type, usually referred to as kinetic equations, suitable to describe the time-evolution of some \emph{social} trait of the agents, like wealth, opinion, knowledge, or others  \cite{CCCC,NPT,ParTos-2013}.

Similarly to kinetic theory of rarefied gases, once inserted into the equation of Boltzmann type, the elementary laws of interactions between agents determine both the time evolution of the density profile and, eventually,  the equilibrium distribution that should resume at best, at least for large times, the statistical characteristics of the phenomenon under investigation. 

Once the emergent steady profile relative to a social phenomenon has been identified, various features allow to have a more precise measurement of its social meaning,  to better understand in this way the macroscopic effect of the microscopic behavioral interactions of agents.

Among the various features that can be introduced to measure the properties of equilibria emerging from kinetic equations that model social phenomena, a relevant importance has been assumed by inequality indices, quantitative scores that in general take values in the unit interval, with the zero score characterizing perfect equality \cite{sinha2019, joseph2022, toscani2022, toscani2025}. 

 Measuring inequalities is a question which is nowadays a priority also in the European agenda \cite{anand2020,alberti2021}.  Indeed, the pursuit of a more equal and fairer Europe requires extensive knowledge on prevailing inequalities across multiple life domains. Against this backdrop, the EU Multidimensional Inequality Monitoring Framework has been recently introduced to contribute to the measurement, monitoring, and analysis of a wide range of different aspects of inequality.\footnote{https://composite-indicators.jrc.ec.europa.eu/multidimensional-inequality}

 In general, the study of inequality in social and economic phenomena is linked to the goal of decreasing inequalities by allowing individuals in a society to have possibly equal opportunities. In particular, in the presence of a mathematical model describing the evolution of the statistical density of wealth in a multi-agent society, it results of high interest to understand if the model drives or not the solution towards a society with a low degree of inequality. An example related to the choice of an optimal taxation strategy to obtain the lowest inequality in the society can be found in Chapter 5 of \cite{ParTos-2013}. There, inequality has been measured in terms of the coefficient of variation.  Also the relationships between the education system and inequalities in the wealth distribution has been discussed in \cite{dimarco2024} in terms of the Gini index \cite{gini1914}.
These results are highly dependent on the existence of a stationary distribution, which allows for explicit computations of the expected degree of inequality provided by the model. 

A more delicate question is to understand how to measure inequalities in a system which is typically characterized by cycles, which prevent the solution to converge towards a steady state. 

In this paper, we will try to give a (partial) answer to this challenging problem, by studying the evolution in time of the coefficient of variation, a simple but efficient measure of inequality \cite{auricchio2024}, in a system of Fokker--Planck equations which has recently been proposed in \cite{TosZan,BMTZ} to study the time evolution of the size distribution of two populations governed by predator--prey interactions. In \cite{TosZan,BMTZ} the evolution in time of the population's sizes has been shown to obey to a system of classical Boltzmann-type equations, where the dynamics arise from elementary binary interactions between the populations. Then, under a suitable scaling regime,  the Boltzmann formulation is close to a system of coupled Fokker--Planck-type equations for the probability densities $f(t,x)$ and $g(t,y)$, with $x,y \in \R_+$ and $t \ge 0$ representing the statistical distributions of preys, and, respectively, of predators.
 The general system found in \cite{BMTZ} reads
 \begin{equations}\label{FP}
\frac{\partial f(t,x)}{\partial t} &= \frac{\partial}{\partial x} \left\{ 
\frac{\sigma m_g(t)}{2} \frac{\partial}{\partial x} \left( x f(t,x) \right)
+ \left[ \left( \beta m_g(t) + \alpha \chi \right) x - \alpha(\chi + 1) m_f(t) \right] f(t,x)
\right\}, \\
\frac{\partial g(t,y)}{\partial t} &= \frac{\partial}{\partial y} \left\{ 
\frac{\tilde{\sigma} m_f(t)}{2} \frac{\partial}{\partial y} \left( y g(t,y) \right)
+ \left[ \gamma \left( \mu - m_f(t) \right) y - \nu m_g(t) \right] g(t,y)
\right\},
\end{equations}
where the positive parameters $\alpha,\beta,\gamma, \mu,\nu,\chi, \sigma, \tilde\sigma$, such that $\gamma\mu-\nu >0$, are linked to the details of the interactions and growth of the two populations, while $m_f(t)$ and $m_g(t)$ represent the mean values of the probability densities $f$ and $g$, i.e.
\be\label{mean}
m_f(t)=\int_{\mathbb{R}_+}x\,f(t,x)\,dx; \quad
    m_g(t)=\int_{\mathbb{R}_+}y\,g(t,y)\,dx.
\ee
Without entering into the detailed description of the meaning of the parameters, which will be presented in the next Section, we outline that the values $\sigma$ and $\tilde\sigma$ provide a measure of the risk associated with the random component of interactions between the two populations.  It is easy to show  that  the mean values $m_f(t)$ and $m_g(t)$ satisfy the Lotka--Volterra system
\begin{equation}\label{LV}
\begin{aligned}
\frac{d}{dt} m_f(t) &= \alpha\, m_f(t) - \beta\, m_f(t)\, m_g(t), \\
\frac{d}{dt} m_g(t) &= -(\gamma\mu-\nu)\, m_g(t) + \gamma\, m_f(t)\, m_g(t),
\end{aligned}
\end{equation}
and that equations \fer{LV} do not depend on the parameters $\sigma$ and $\tilde\sigma$. Consequently, while
 the system of Fokker--Planck equations \fer{FP} follows the evolution of the probability densities and links the macroscopic dynamics of their mean values to the Lotka--Volterra system of ordinary differential equations \fer{LV}, it contains more detailed information with respect to it, which only gives details on the evolution of the first moments of the probability densities $f$ and $g$. Furthermore, the system of Fokker--Planck equations possesses a pair of local equilibria, in the form of Gamma densities with variable in time coefficients \cite{BMTZ}, which in principle could be useful to extract information about the large--time behavior of the densities of the two populations.

The Fokker--Planck system studied in \cite{BMTZ} allows for economic and social interpretations, which are consistently obtained by resorting to the numerous applications to economy of Lotka--Volterra equations present in the literature, starting from the pioneering work of Palomba \cite{Palomba1939}, the contribution of Goodwin \cite{goodwin1955}, Solomon and coworkers \cite{Solomon-2002, Malcai-2002}, and ending with the recent contribution of 
Sumarti,  Nurfitriyana, and Nurwenda \cite{SNN-2014} in which Lotka--Volterra-type equations have been introduced to describe a dynamical system of deposit and loan volumes of a bank, where the predators are loan volumes, and the prey are deposit volumes.

In addition, Lotka--Volterra equations have been interpreted within a social framework \cite{chakra2016} to describe a pattern of technological evolution due to the interaction between multiple countries and the resulting effects on the corresponding macro variables. In this framework, the world consists of a set of economies in which some countries are leaders and others followers on the technological ladder. 

Owing to the previous discussion, it is clear that the evolution of the solutions of the Fokker--Planck system \fer{FP} can be fruitfully interpreted in any of the previous economic and social contexts, characterized by a precise economic and/or social meaning of the pair of variables $x,y$. Consequently, the study of the evolution in time of inequality indexes applied to the pair of probability densities $f$ and $g$, solutions of the Fokker--Planck system \fer{FP} which are generally assumed of finite mean and variance, constitutes a primary issue.

However, to our knowledge, this kind of problem has not been much studied in the literature.
For this reason, in this paper we will present a possible approach to measure inequality in system \fer{FP} of Fokker--Planck-type equations in the case in which they are assumed to describe the economical situation postulated in \cite{SNN-2014}. 

Our main argument will be the study of the time evolution of the coefficient of variations of
the probability densities solutions of \fer{FP}, in two different situations, characterized by the different levels of risk assumed for the loan volumes of banks. In particular, in presence of a high level of risk, the analysis of the time--behavior of the coefficient of variation of the distribution of loans operated by banks allows to  conclude that the corresponding quasi--equilibrium density do not furnish in general a precise picture of the real behavior of the inequality present in the system.   

In more details, in Section \ref{sec:Gini} we will recall some information about the coefficient of variation and the well-known Gini index, by enlightening some useful connections between them.  Then, in Section \ref{sec:model} we will introduce and discuss the deposit--loan kinetic model, first in terms of the elementary interactions, and subsequently by its kinetic description in terms of the Fokker--Planck system \fer{FP}. Sections \ref{sec:CV} and Section \ref{sec:risk} will be devoted to the study of the evolution of the coefficients of variation of the probability densities which solve system \fer{FP} in the two cases in which the loans of the banks are subject to moderate and/or high risk. Last, Section \ref{sec:num} contains some numerical experiments that help to clarify the large-time behavior of the inequality of the solutions to the system.  


\section{Basic facts about the coefficient of variation}\label{sec:Gini}

\subsection{Gini index and the coefficient of variation}

The coefficient of variation is the standard measure to summarize through a scalar value the dispersion of a set of points in a statistical distribution.
For one-dimensional distributions, the universally accepted definition of the coefficient of variation  is: 
\begin{equation}
    \label{eq:std_CV}
    c_f=\frac{\sigma}{|m|},
\end{equation}
where $\sigma^2$ and $m$ are the variance and mean, respectively, of a probability density $f$.

The Gini index, in its univariate form, is a measure of dispersion that quantifies the spread of a distribution using a single scalar value. It reflects the mean difference between the values of the distribution, normalized by its mean.
Given a probability density $f$ over $\mathbb{R}$ with a nonzero mean $m\neq 0$, the Gini index \cite{gini1914,gini1921measurement} is formally defined as: 
\begin{equation} \label{eq:Gini_intro} G(f):=\frac{1}{2|m|}\int_{\mathbb{R}}\int_{\mathbb{R}}|x-y|f(x)f(y)\, dxdy. 
\end{equation}
Notably, when $f$ is supported on the interval $[0,\infty)$, the index satisfies $0\leq G(f)\leq 1$ \cite{gini1914,gini1921measurement}, making its interpretation more intuitive.
In fact, if $G(f)=0$, then $f$ is a Dirac delta distribution, which means that it is a probability measure concentrated entirely at a single point. This represents a state of perfect equality, as every realization of the associated random variable yields the same value.
Conversely, as $G(f)$ approaches 1, the associated random variable exhibits greater dispersion.

Furthermore, while the coefficient of variation can be defined for any probability distribution with a non-zero mean, the Gini index requires $f$ to be supported exclusively on the positive half line to ensure it remains within the range of $0$ to $1$.

Despite their differences, the coefficient of variation in \eqref{eq:std_CV} and the Gini index in \eqref{eq:Gini_intro} share several key characteristics. Both serve as measures of dispersion, summarizing a distribution’s spread through a scalar value that remains independent of the unit of measurement.
These similarities become particularly evident when examining a Gaussian distribution.
In fact, if $f$ is a Gaussian distribution of mean $m$ and variance $\sigma^2$, a simple computation shows that its Gini index is equal to  
\[
G(f) = \frac 2{\sqrt\pi}\frac{\sigma}{|m|}.
\]
Thus, for a Gaussian distribution, the Gini index is proportional to the coefficient of variation.
Furthermore, if we consider an alternative \emph{squared} definition of the Gini index \cite{auricchio2024}:
\begin{equation}
\label{eq:G2_intro}
    G_2(f):=\Big(\frac{1}{2m^2}\int_{\erre}\int_{\erre}|x-y|^2f(x)f(y)\, dxdy\Big)^{\frac{1}{2}},
\end{equation}
based on the $L^2$ norm rather than the $L^1$ norm as in \eqref{eq:Gini_intro}, we have that
\[
    G_2(f)=\frac{\sigma}{|m|},
\]
which shows that, for any distribution $f$, the squared Gini index $G_2$ coincides exactly with the coefficient of variation.
The above findings indicate that the Gini index $G_1$ and the coefficient of variation are the $L^1$ and $L^2$ norms of the same function ${|x-y|}/{|m|}$.
\subsection{An example from wealth distribution}
A simple example that clarifies the role that the coefficient of variation can play in applications directed at measuring the dispersion of the solution to Fokker--Planck type equations, is concerned with the Fokker--Planck equation introduced by Bouchaud and M{\'e}zard in \cite{BM-2000} to study the evolution of wealth in a multi-agent society \cite{ParTos-2013}. This equation, which reads
  \be\label{Wealth}
 \frac{\partial h}{\partial t} = \frac \sigma{2}\frac{\partial^2 }
 {\partial x^2}\left( x^2 h\right) + \frac{\partial }{\partial x}\left(
 (x-m) h\right),
 \ee
 describes the evolution of the wealth distribution density $h(x,t)$  towards a steady state.\\
In \fer{Wealth}  $m$  and $\sigma$  denote two positive constants, with $\sigma$ quantifying the risk.  Equation \fer{Wealth}  was subsequently obtained in \cite{CoPaTo05} via an asymptotic procedure from a Boltzmann-type kinetic model for trading agents. \\
 The unique stationary solution of unit mass of \fer{Wealth} is given by the inverse
 Gamma distribution \cite{BM-2000,CoPaTo05}
 \be\label{equi2}
h_\infty(x) =\frac{(\mu-1)^\mu}{\Gamma(\mu)}\frac 1{x^{1+\mu}}{\exp\left\{-\frac{m(\mu-1)}{x}\right\}},
 \ee
where
  $$ \mu = 1 +  \frac{2}{\sigma} >1.
$$
This stationary distribution exhibits a power-law tail for large values of the wealth variable, and, in particular possesses finite variance if and only if $\sigma < 2$.

Let us fix $\sigma <2$. It is immediate to show that, if like in \fer{mean}, $m_h(t)$ denotes the mean value of the solution $h(x,t)$ of equation \fer{Wealth}, the square of the coefficient of variation $c_h(t)$ satisfies the first-order linear differential equation
\be\label{ch}
\frac{d}{dt} c_h(t)^2= -\left( 2\frac m{m_h(t)}-\sigma\right)c_h(t)^2 + 2\frac m{m_h(t)}.
\ee
Hence, the evolution of $c_h(t)$ depends on the evolution of the mean value $m_h(t)$, which is explicitly known. Indeed
\[
m_h(t) -m = m_h(0) + (m_h(0) -m) e^{-t}.
\]
In the particular case in which $m_h(0) =m$, $m_h(t) = m$ for all times, and the solution to \fer{ch} is explicitly given by
\[
c_h(t) = \left[ \frac 2{2-\sigma} + \left(c_h(0)^2 - \frac 2{2-\sigma}\right) \exp{\{-(2-\sigma)\}} \right]^{1/2},
\]
and this implies exponential convergence of $c_h(t) $ towards the value $\sqrt{2/(2-\sigma)}$.

In this case, the equilibrium distribution is an inverse Gamma distribution. The Gini index of an inverse Gamma distribution of shape parameter $a>2$ and rate parameter $\lambda>0$
\[
f(x) = \frac{\lambda^a}{\Gamma(a)}x^{-(a+1)}e^{-\lambda/ x},
\]
is equal to \cite{Gamma-Gini}
\[
G(f) = \frac{\Gamma\left(a-\frac 12 \right)}{\sqrt\pi\,\Gamma(a)},
\]
while its coefficient of variation is
\[
c_f = \frac 1{\sqrt{a-2}}
\]
Now, Gautschi's inequality for Gamma function says that,  for any positive real number $a$  it holds
\[
(a-1)^{1/2} < \frac{\Gamma(a)}{\Gamma\left(a-\frac 12 \right)} < a^{1/2}. 
\]
Therefore, for a given inverse Gamma distribution $f$ of shape parameter $a$ 
\be\label{relG}
 c_f \sqrt{\frac{a-2}a}  < \sqrt\pi G(f) < c_f \sqrt{\frac{a-1}a},
\ee
hence an explicit relationship between the standard Gini index and the coefficient of variation of a Gamma distribution, which justifies in this case the choice of the study of the evolution of the latter.

\section{An economic application of the kinetic Lotka--Volterra system}\label{sec:model}

\subsection{The deposit-loan model}

In this section, according to the proposal in \cite{SNN-2014},  we will briefly show that system \fer{FP} can be used to obtain a statistical description of the joint time-evolution of the distribution of the amount $x$ of money placed in bank deposits, say  $f(x,t)$, by a population of agents,  and the distribution of loans of amount $y$ operated by banks, say $g(x,t)$. 

The interested reader can find a detailed explanation about microeconomics of banking in the recent books \cite{FR-2008,MT-2023}. 
In short, a bank's balance sheet consists of three key components: liabilities, assets, and equity. Liabilities represent obligations that the bank must fulfill in the future, such as repayments to depositors. Assets are economic resources expected to generate future benefits. Equity reflects the bank's net worth, calculated as the difference between total assets and total liabilities.
Typically, a bank's balance sheet follows the equation: Total Assets = Total Liabilities + Equity.
This equation underscores the relationship between these three elements, forming the foundation of financial accounting in the banking sector.

The main liability of a bank is deposits, and assets of a bank are including Loan and Reserves (primary and secondary).
In a simplifying model, one has 
\[ 
L+R_1+R_2 = D 
\] 
where $R_1$ and $R_2$ are the primary and secondary reserves, respectively. All variables in the assets are proportions of the volume of deposit. It can be easily realized that the existence of loan and both reserves depend on the volume of deposits, so that part of bank activities can be seen as a predator-prey interaction, where the loan and the reserves are predators and the deposit are prey. 

According to this predator-prey description, as proposed in \cite{SNN-2014}, the Lotka--Volterra sytem \fer{LV} can be adapted to describe the dynamics of the mean value $m_f(t)$ of deposit, and, respectively, the mean value $m_g(t)$ of loan volumes. In this picture, the parameter $\alpha$ in \fer{LV} measures the interest rate of deposit where the increase of its value will lead to the increase of deposit volume. The parameter $\gamma\mu -\nu$ measures the interest rate of loans, where an increase in its value will lead to a decrease in loan volume. Moreover in \fer{LV} it is assumed that $\beta =\gamma$, so that the rate of deposit volume with respect to time will decrease by $\gamma m_f(t)m_g(t)$, the bank’s decision on the mixture between existing volumes of deposit and loan volume, while the loan volume will increase of the same amount. In what follows, to remain as general as possible, we will assume the possibility that the parameters $\beta$ and $\gamma$ are different.

\subsection{The kinetic description}

We follow, hereafter, the kinetic approach developed in \cite{BMTZ}, where the interested reader is referred to for details. Let $f(t,x)$ and $g(t,y)$, for $t \geq0$, be the distribution functions of deposits operated by the population of agents, and, respectively, of loans granted by banks, measured in terms of a certain currency. The evolution in time of $f(t,x)$ and $g(t,y)$, which are initially assumed to be probability density functions, so that
\begin{equation*}
\int_{\R_+} f(x,0)\, \dd x = \int_{\R_+} g(y,0)\, \dd y = 1,
\end{equation*}
is driven by two different processes. The first one is referred to the elementary interactions between the two populations, which depend on both deterministic and random effects. These interactions read 
\begin{equation} \label{eqmicro}
    \begin{split}
        x' &= x - \Phi(y)x + \left(\frac{x}{\bar{x}}\right)^p \mathds{1}(x \geq (1-p)s_0) \eta_1(y),
        \\[2mm]
        y' &= y + \Psi(x)y + \left(\frac{y}{\bar{y}}\right)^p \mathds{1}(y \geq (1-p)s_0) \eta_2(x).
    \end{split}
\end{equation}
In particular, the deterministic contributions 
\begin{equation*}
\Phi(y) = \beta \frac{y}{1 + y}  \quad 
\Psi(x) = \gamma \frac{x - \mu}{1 + x} \, 
\end{equation*}
quantify the amount of money employed in deposits or loans, in which the amount of agent's deposit is chosen to depend on the financial consistency of the bank, and, on the other size, the size of loans depend on the financial consistency of the agent. As in \cite{BMTZ}, the contributions 
are Holling-type II functional responses,  where $\beta, \gamma \in (0,1)$ and $\mu \geq 1$ are positive constants. Hereafter, we assume that
\begin{equation}\label{gm}
    \gamma \mu < 1.
\end{equation}
The random contributions, which in this case quantify both the possible gain and/or the possible risks, are given by  $\eta_f$ and $\eta_g$,  independent random variables of zero mean and bounded variance, given by
\begin{equation}
\langle \eta_f^2(y) \rangle = \sigma_f \frac{y}{1 + y} \, ; \quad 
\langle \eta_2^2(x) \rangle = \sigma_g \frac{x}{1 + x} \, .
\end{equation}
The quantity $\langle \cdot \rangle$ represents the expected value with respect to, respectively, $\eta_f$ and $\eta_g$, and $\sigma_f, \sigma_g$ are positive parameters. Finally, $\mathds{1}(A)$ is the characteristic function of the set $A \subseteq \R_+$, which is needed to guarantee the positivity of the post-interactions sizes $x'$ and $y'$. Last, the value $p$, with $0<p\le 1$ quantifies the degree of risk. Values of $p< 1$ correspond to a low risk, while $p=1$ indicates a high risk proportional to the amount of money put into the financial operation. In \cite{BMTZ}, the cases $p=1/2$ (medium risk) and $p =1$ (high risk), were considered, in view of the possibility of performing explicit computations.

The second process is characterized by suitable redistribution processes describing both the positive return of money consequent to deposits of the population and to the negative return of money for banks consequent to their exit due to loans. Following \cite{BMTZ}, denoting by $z \in \R_+$  the variable representing the size of resources, the redistribution operator acts at the particle level as
\begin{equation} \label{eqmicro2}
    \begin{split}
        x'' &= x + \alpha \left(z - \chi x \right), \\[2mm]
        y'' &= y + \nu \left(z - \theta y \right),
    \end{split}
\end{equation}
where $\alpha, \nu > 0$ are constant interaction coefficients, while $\chi, \theta > -1$ define constant rates of variation for the two populations. In particular
\begin{equation}
    \alpha\chi < 1, \quad \nu\theta < 1.
\end{equation}
\noindent Bearing the microscopic interactions \eqref{eqmicro} and \eqref{eqmicro2} in mind, at the mesoscopic level the evolution of the system is provided by the following Boltmann-type system, \cite{ParTos-2013, TosZan}:

\begin{equation} \label{eq:Boltzmann}
    \begin{split}
        \displaystyle \frac{\partial f(x,t)}{\partial t} = R_{\chi}^{\alpha}(f_1)(x,t)+Q_{12}(f_1,f_2)(x,t), \\[2mm] 
        \displaystyle \frac{\partial g(y,t)}{\partial t} = R_{\theta}^{\nu}(f_2)(y,t)+Q_{21}(f_2,f_1)(y,t),
    \end{split}
\end{equation}
where $R_\chi^\alpha(f_1)$ and $R_\theta^\nu(f_2)$ are redistribution-type linear operators describing the balance of the microscopic interactions \eqref{eqmicro}, whereas $Q_{12}(f_1,f_2)$ and $Q_{21}(f_2,f_1)$ are Boltzmann-type bilinear operators accounting for the microscopic interactions defined by \eqref{eqmicro2}.




In a regime characterized by an increasing number of elementary interactions of small size (quesi-invariant regime), it has been shown in \cite{BMTZ} that the Boltzmann system   \eqref{eq:Boltzmann} is well described by system \fer{FP}, a system of Fokker--Planck type equations, which are characterized by variable in time coefficients of diffusion and drift. 
It is important to remark that the conservation of mass requires that system \fer{FP} is complemented with no-flux boundary conditions at $x, y = 0$.

In the present situation, the system of Lotka--Volterra type describes the evolution of the distribution densities of deposits and loans.
As shown in \cite{BMTZ},  the evolution of the mean values \fer{mean}
satisfies the classical Lotka--Volterra model \fer{LV}, so that for any initial data $m_f(0)$, $m_g(0) > 0$, a unique vector solution $\mathbf{m}(t) = (m_f(t), m_g(t))$ exists for all times and remains positive. It is important to observe that the Lotka--Volterra system does not depend on the random part of the elementary interaction \fer{eqmicro}. 

However, at variance with the classical description of the predator--prey interaction, the Fokker--Planck description of system \fer{FP} allows to compute also the evolution of higher moments of the distributions. In particular, it is immediate to express the time-evolution  of the variances 
of the distributions $f$ and $g$, given by 
\begin{align}\label{var}
    v_f(t)=\int_{\mathbb{R}_+}(x-m_f(t))^2f(t,x)\,dx\quad v_f(t)=\int_{\mathbb{R}_+}(x-m_f(t))^2f(t,x)\,dx.
\end{align}
If we choose a risk parameter $p =1/2$, the evolution of the variances is given by the system
\begin{equation} \label{eq:variance FP p=1/2}
    \begin{split}
        & \frac{\dd v_f(t)}{\dd t} = -2(\beta m_g(t) + \alpha\chi) v_f(t) + \sigma_f m_f(t) m_g(t), \\[2mm]
        & \frac{\dd v_g(t)}{\dd t} = -2 \big( \gamma(\mu - m_f(t)) + \nu\theta \big) v_g(t) + \sigma_g m_f(t) m_g(t),
    \end{split}
\end{equation}
which clearly shows that the variation in time of the size of the variances is directly proportional to the variances $\sigma_f$ and $\sigma_g$ of the random variables quantifying the risk. 

In addition, the formulation of a prey--predator system in terms of a system of Fokker--Planck type equations, allows for the possibility to explicitly obtain the shape of the local equilibrium states (or \emph{quasi-equilibria}), namely of the distribution densities which satisfy system \fer{FP} with the left--hand side equal to zero. Denoting with $\mathbf{f}^\eq(\x,t) = (f_1^\eq(x,t),f_2^\eq(y,t))$ the vector of quasi--equilibria, in the case $p =1/2$ we obtain
\begin{equation} \label{eq:equilibrium states}
    \begin{split}
        f^\eq(x,t) & = C_f(t) x^{\frac{2\alpha(\chi+1)}{\sigma_f}\frac{m_f(t)}{m_g(t)} - 1} \exp\left\{-\frac{2}{\sigma_f m_g(t)} (\beta m_g(t) + \alpha \chi) x \right\},
        \\[2mm]
        g^\eq(y,t) & = C_g(t) y^{\frac{2 \nu(\theta+1)}{\sigma_g}\frac{m_g(t)}{m_f(t)} - 1} \exp\left\{-\frac{2}{\sigma_g m_f(t)} (\gamma(\mu - m_f(t) + \nu\theta) y \right\},
    \end{split}
\end{equation}
where $C_f(t)$, $C_g(t) > 0$ are normalization coefficients depending on the means $m_f(t)$ and $m_g(t)$. It is worth noting that the equilibrium relative to a low risk takes the form of a Gamma distribution.


\section{Towards a study of the inequality of the deposit-loan system}\label{sec:CV}

As briefly shown in Section \ref{sec:Gini}, a simple way to quantify the evolution in time of the inequality linked to the probability densities of the two interacting populations is to resort to the analysis of the variable-in-time coefficients of variation of the two populations, defined by
\begin{equation}\label{CV}
    c_f(t)=\frac{\sqrt{v_f(t)}}{m_f(t)}, \qquad
    c_g(t)=\frac{\sqrt{v_g(t)}}{m_g(t)}.
\end{equation}
Notice that, since the mean values of the Lotka-Volterra system \fer{LV} are always bounded from below by a positive constant, the quantities in \fer{CV} are always well-defined.

It is immediate to show that,  owing to system \eqref{eq:variance FP p=1/2}, the variations in time of the coefficients in \fer{CV} satisfy the system
\begin{equation}\label{eqsysfp2}
    \begin{cases}
    \displaystyle\frac{d}{dt}c_f(t)&=-\alpha(\chi+1) c_f(t)+\displaystyle\frac{\sigma_f}{2}\frac{m_g(t)}{m_f(t)}\frac 1{c_f(t)}\\\\
   \displaystyle \frac{d}{dt}c_g(t)&=-\nu (\theta+1)c_g(t) +\displaystyle\frac{\sigma_g}{2}\frac{m_f(t)}{m_g(t)}\frac 1{c_g(t)}.
    \end{cases}
\end{equation}
System \fer{CV} is equivalent to
\begin{equation}\label{CV2}
    \begin{cases}
    \displaystyle\frac{d}{dt}c_f^2(t)&=-2\alpha(\chi+1) c_f^2(t)+\displaystyle{\sigma_f}\frac{m_g(t)}{m_f(t)}\\\\
   \displaystyle \frac{d}{dt}c_g^2(t)&=-2\nu (\theta+1)c_g^2(t) +\displaystyle{\sigma_g}\frac{m_f(t)}{m_g(t)}.
    \end{cases}
\end{equation}
It is interesting to remark that, like in the simple example of Section \ref{sec:Gini}, equation \fer{ch}, the evolution of the squares of the coefficients of variation satisfy two first-order linear differential equations. In addition the evolution of the coefficients of variation only depends on the mean values of the two populations through their time-dependent ratios. In particular, the influence of the ratios on the evolution depends of the degree of randomness (the risk)  present in the system of the Fokker--Planck equations through the coefficients $\sigma_f$ and $\sigma_g$ quantifying their diffusion terms.
Indeed, in absence of randomness, the coefficients of variation collapse exponentially towards the value zero, hence towards the state of perfect equality. In other words, inequality in time into the population densities is maintained by the presence of randomness in the process of interaction.
A second aspect to be remarked is that the intensity of the redistribution processes, characterized by the positive quantities $\chi$ and $\theta$ plays against inequality, since it works in favor of the increase of the speed of decay of the coefficients.

Given the initial values of the means and of the coefficients of variation, expressed by
\begin{align*}
   m_f(0),\quad m_g(0),\quad c_f(0),\quad c_g(0), 
\end{align*}
integration of \fer{CV2} gives
\begin{equation}\label{eqboundcv}\begin{split}
 \displaystyle  c_f(t)&=\left[c_f^2(0) e^{-2 \alpha(\chi+1) t} + \sigma_f e^{-2 \alpha(\chi+1) t} \int_0^t \frac{m_g(s)}{m_f(s)} e^{2 \alpha(\chi+1) s} \dd s,\right]^{1/2},\\[2mm]
    \displaystyle c_g(t)&=\left[ c^2_g(0) e^{-2  \nu(\theta+1) t} + \sigma_g  e^{-2 \nu(\theta+1) t} \int_0^t \frac{m_f(s)}{m_g(s)} e^{2 \nu(\theta+1) s}\dd s\right]^{1/2}  . 
\end{split}\end{equation}

Therefore, for large times, the coefficients of variation in \fer{eqboundcv} tend to lose dependence on their initial values, and, if 
\be\label{bbb}
0< r\le \frac{m_g(t)}{m_f(t)} \le R <+\infty,
\ee
a simple application of the Hopital's rule shows that for large times $c_f(t)$  oscillates between the two values  
\[
\left(\frac{\sigma_f}{2\alpha(\chi +1)}r \right)^{1/2}\le c_f(t) \le \left(\frac{\sigma_f}{2\alpha(\chi +1)}R\right)^{1/2}.
\]
Similar result for $c_g(t)$.
It is interesting to compare the large--time bounds of the coefficients of variations \fer{eqboundcv} of the solutions to the Fokker--Planck system \fer{FP} with the coefficients of variations of the quasi--equilibria \fer{eq:equilibrium states}. Since these distributions are Gamma densities, their coefficients of variation are easily computed from the knowledge of the mean value and variance of this type of density. One obtains
\begin{equation}
  c_{f^{eq}}(t) = \left(\frac{\sigma_f}{2\alpha(\chi +1)}\frac{m_g(t)}{m_f(t)} \right)^{1/2} ; \qquad
   c_{g^{eq}}(t) = \left(\frac{\sigma_g}{2\nu(\theta +1)}\frac{m_f(t)}{m_g(t)} \right)^{1/2}.
\end{equation}
Therefore, the coefficients of variations of the quasi-equilibria describe well the large--time behavior of the coefficients of variation of the true solutions.


\section{The importance of risk factors}\label{sec:risk}

The clean evolution of the coefficients of variation in the economic example of the kinetic Lotka--Volterra system is mainly due to the choice of a risk factor characterized by $p=1/2$ in \fer{eqmicro}. To better follow the consequences of this choice, let us assume that the two groups are characterized by different levels of risk. For example, let us assume that the population of agents is characterized by a lower risk ($p=1/2$), whereas the banks are characterized by a higher risk ($p=1$). 
It is worth pointing out that these two cases have already been analyzed in \cite{TosZan, BMTZ}, separately. 
At variance with system \fer{FP}, the Fokker-Planck equation for banks (the interested reader is referred to \cite{TosZan,BMTZ} for technical details) is now of the form
\begin{equation} \label{eq:Fokker-Planck2}
     \frac{\partial g(y,t)}{\partial t} = \frac{\sigma_gm_f(t)}{2} \frac{\partial^2}{\partial y^2} \left[y^{2}g(y,t)\right] + \frac{\partial}{\partial y} \big( (\gamma(\mu - m_f(t)) y + \nu\theta y -\nu(\theta+1) m_g(t)) g(y,t) \big).
  \end{equation}
At the macroscopic level, the system of mean values still remains   \eqref{LV}. 
However, the system of variances modifies into
\begin{equation}\label{eqvarnew}
\begin{aligned}
    \frac{d}{dt}v_f(t) &= -2(\beta m_g(t) + \alpha\chi) v_f(t) + \sigma_f m_f(t) m_g(t)\\
    \frac{d}{dt}v_g(t)&=-2\left(\gamma\left(\mu-\left(1-\frac{\sigma_f}{2\gamma}\right)m_f(t)\right)+\nu \theta\right)v_g+\sigma_g m_f(t) m_g^2(t).
\end{aligned}
\end{equation}
 Now, let us compute the evolution of the coefficients of variation $c_f(t)$ and $c_g(t)$ by using the system of means \eqref{LV} and the system of variances \eqref{eqvarnew}. We obtain
\begin{equation}\label{CV3}\begin{cases}
    \displaystyle\frac{d}{dt}c_f(t)=-\alpha (\chi+1)c_f(t)+\frac{\sigma_f}{2}\frac{m_g(t)}{m_f(t)}\frac{1}{c_f(t)}\\\\
    \displaystyle\frac{d}{dt}c_g(t)=-\left[\nu(\theta+1)+\frac{\sigma_g}{2}m_f(t)\right]c_g(t)+\frac{\sigma_g}{2}\frac{m_f(t)}{c_g(t)}, 
\end{cases}\end{equation}
that, equivalently, can be expressed as
\begin{equation}\label{CV4}
\begin{cases}
    \displaystyle\frac{d}{dt}c_f^2(t)=-2\alpha (\chi+1)c_f^2(t)+\sigma_f\frac{m_g(t)}{m_f(t)}\\\\
    \displaystyle\frac{d}{dt}c_g^2(t)=-\left[2\nu(\theta+1)+\sigma_g m_f(t)\right]c_g^2(t)+
    \sigma_g
    m_f(t).
\end{cases}\end{equation}
The first equation of system \eqref{CV3}, coincides with the first equation of system \eqref{eqsysfp2}, referred to the value $p={1}/{2}$. The second equation in \eqref{CV4}, relative to a higher risk parameter takes a new form, in which the main difference is relative to the fact that, in addition to the various constant parameters, the evolution of the coefficient of variation $c_g(t)$ depends only on the average value of deposits.  
Moreover, in contrast with the evolution of $c_f(t)$, both coefficients in the evolution of $c_g(t)$ depend on time through the mean value $m_f(t)$. 

It is interesting to remark that, in case the value of the coefficient of variation at time $t=0$ satisfies $c_g(t=0)  \ge 1$, its value in time is monotonically decreasing up to the time $t^*$ in which it reaches the value
\[
c_g(t^*) = \sqrt{\frac{\sigma_g m_f(t^*)}{2\nu(\theta+1) +\sigma_g m_f(t^*)}} < 1.
\]
Hence, a consequence of the presence of high risk is to avoid dispersion. This fact is confirmed by the nature of the quasi-equilibrium, which for $p=1$ takes the form of an inverse Gamma function \cite{BMTZ}
\begin{equation}\label{quasieq}
    \displaystyle g^{\eq}(y,t)  = C_g(t) y^{- p(t)} \exp\left\{-\frac{2 \nu (\theta+1) m_g(t)}{\sigma_g m_f(t)} \frac{1}{y} \right\},
\end{equation}
which is characterized by the variable in time Pareto index 
\[
p(t) = \frac{2\gamma(\mu -1) +2\theta\nu +  \sigma_g m_f(t) }{\sigma_g m_f(t)} + 1
\]
It is important to remark that, since $\mu \ge 1$, Pareto index $p(t) >2$, which ensures that the mean value of the inverse Gamma distribution is bounded. On the other hand, the second moment of the quasi--equilibrium distribution \fer{quasieq} is bounded only if $p(t) > 3$, which holds true in time only when
\[
\sigma_g\sup m_f(t) < 2\gamma(\mu -1) + \theta \nu,
\]
that in general it is not granted. Consequently, in presence of high risk, it can happen that the quasi--equilibrium distribution, at variance with the case of low risk treated in Section \ref{sec:CV}, does not give precise indications about the behavior of the kinetic Lotka--Volterra system for large times.

\section{Numerical experiments}\label{sec:num}

In this Section, we provide some numerical simulations of the deposit-loan system discussed and analyzed in the previous Sections. Specifically, in addition to the numerical experiments developed in \cite{TosZan, BMTZ}, we focus on the evolution in time of the coefficients of variation $c_f(t)$ and $c_g(t)$ of the two populations. In what follows, according to the theoretical approach of  Sections \ref{sec:CV} and \ref{sec:risk}, we focus on both situations $p=1/2$ and $p=1$ for the bank system. For both cases, we refer to the values of parameters defined in \cite{TosZan}, that are collected into Table \ref{tab:parameters}. In addition, in all cases we assume the same initial conditions $(m_f(0), m_g(0), c_f(0), c_g(0))$.
\begin{table}[h!]
\begin{center}
\begin{tabular}{|c|c|c|}
\hline
Parameter		&	Value		&	Meaning \\[1mm]
\hline \hline
$\alpha$		&	1			&	Deposits growth rate \\[1mm]
$\beta$			&	0.5			&	Deposits predation rate \\[1mm]
$\mu$			& 	10			&	Deposits lowest size \\[1mm]
$\gamma$		&	0.15		&	Loans growth rate \\[1mm]
$\nu$			&	1			&	Loans take-up rate \\[1mm]
$\delta$		& $\gamma\mu - \nu$ &	Loans shutdown rate \\[1mm]
$\sigma_1$		&	$10^{-3}$	&	Deposit randomness coefficient \\[1mm]
$\sigma_2$&	$10^{-3}$	&	Loans randomness coefficient \\[1mm]
$\chi$			&	0.8			&	Deposits redistribution
                                    coefficient \\[1mm]
$\theta$		&	0.4			&	Loans redistribution
                                    coefficient \\[1mm]
\hline
\end{tabular}
\end{center}
\caption{Parameters used to solve the coupled differential systems of the means \eqref{LV} and the coefficients of variation \fer{CV}.}
\label{tab:parameters}
\end{table}

\begin{figure}[htbp]
  \centering
  \includegraphics[width=0.45\textwidth]{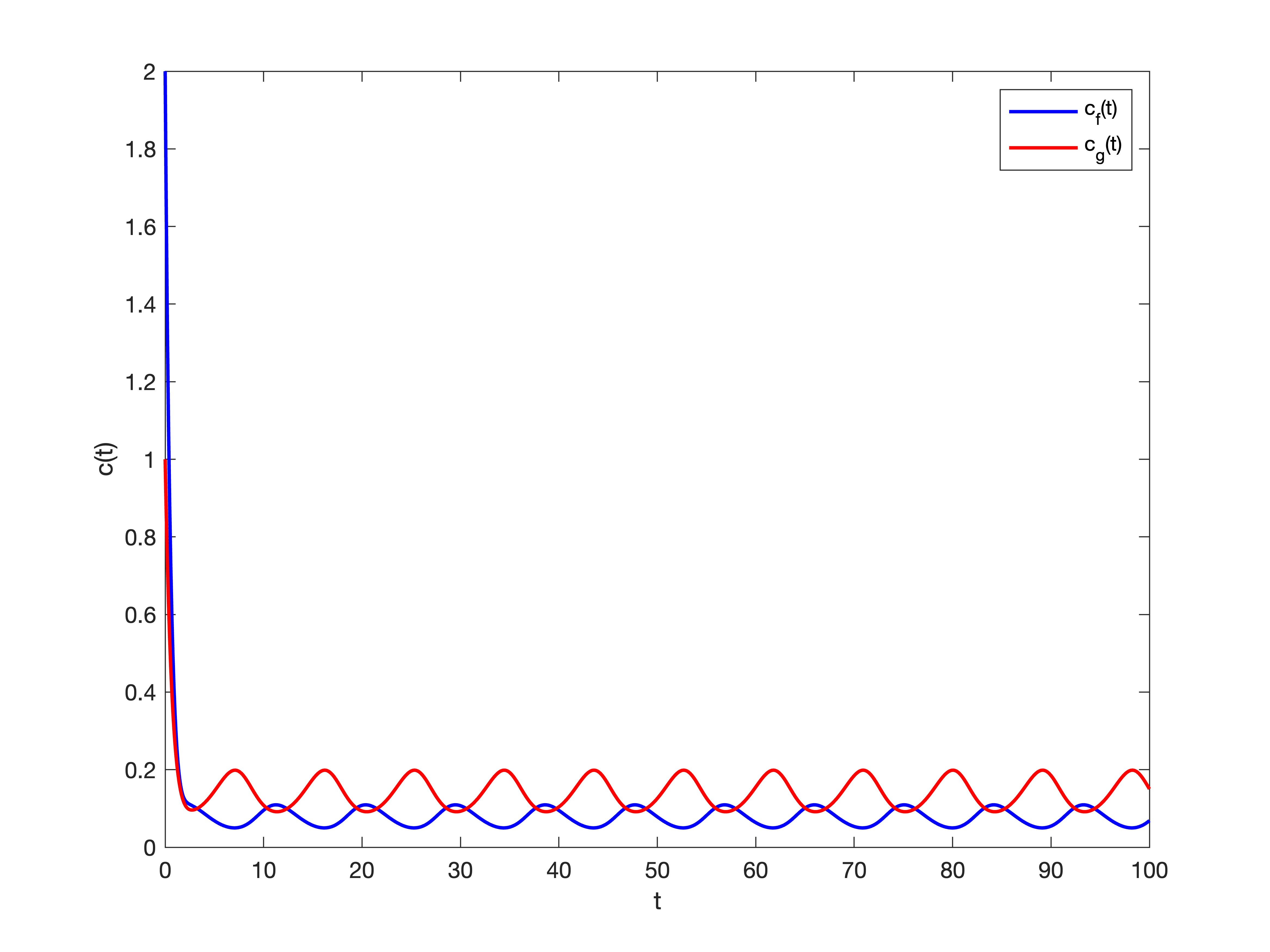}
  \includegraphics[width=0.45\textwidth]{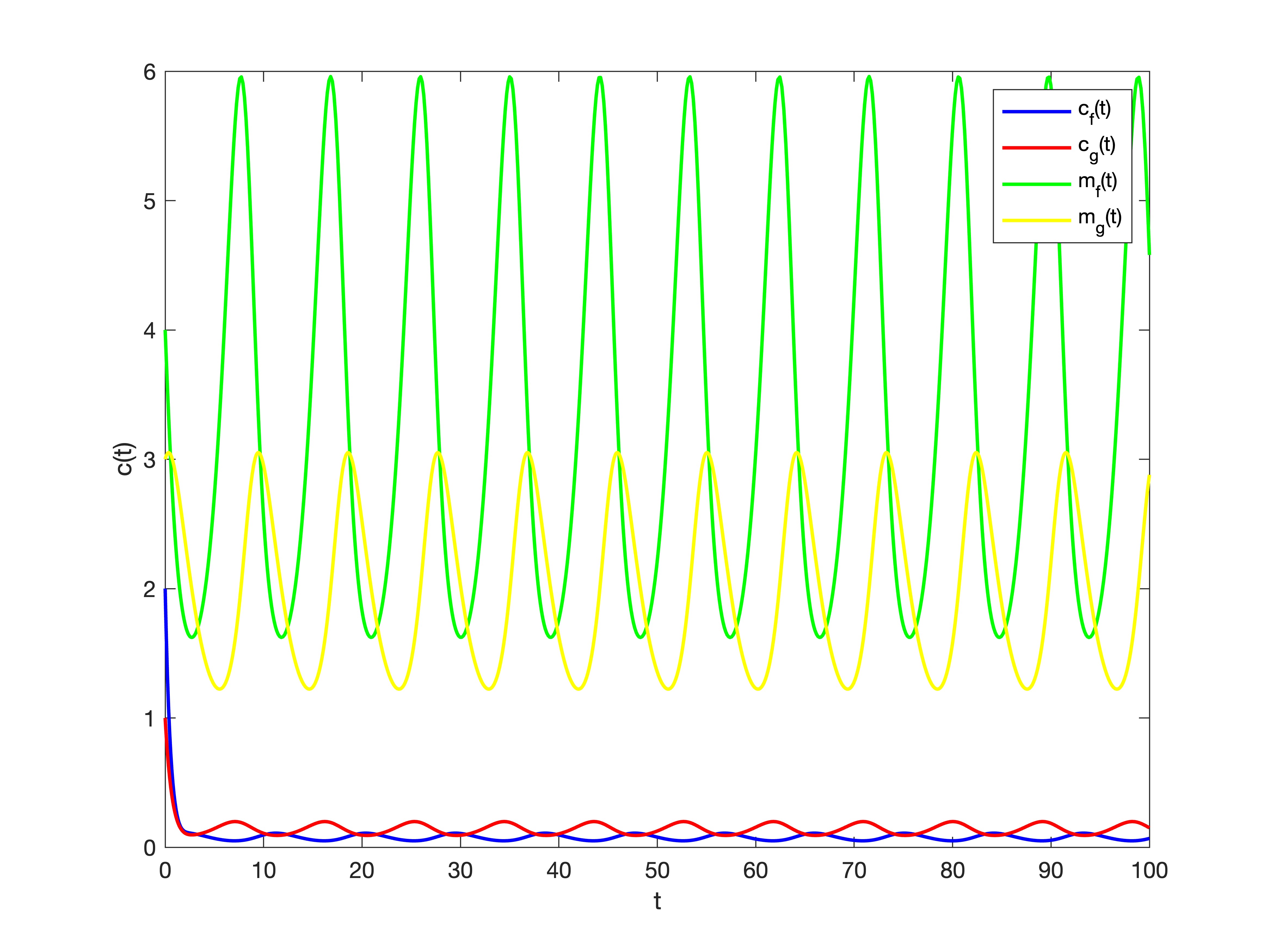}
  \caption{Evolution of the system for $p=\frac{1}{2}$ in both populations. Panel on the left: evolution of coefficients of variation, $c_f(t)$ and $c_g(t)$. Panel on the right: evolution of coefficients of variation, $c_f(t)$ and $c_g(t)$, along with mean values, $m_f(t)$ and $m_g(t)$. Initial data $(m_f(0), m_g(0), c_f(0), c_g(0))=(4,\, 3,\, 2,\, 1)$.}
  \label{fig1}
\end{figure}

\begin{figure}[htbp]
  \centering
  \includegraphics[width=0.45\textwidth]{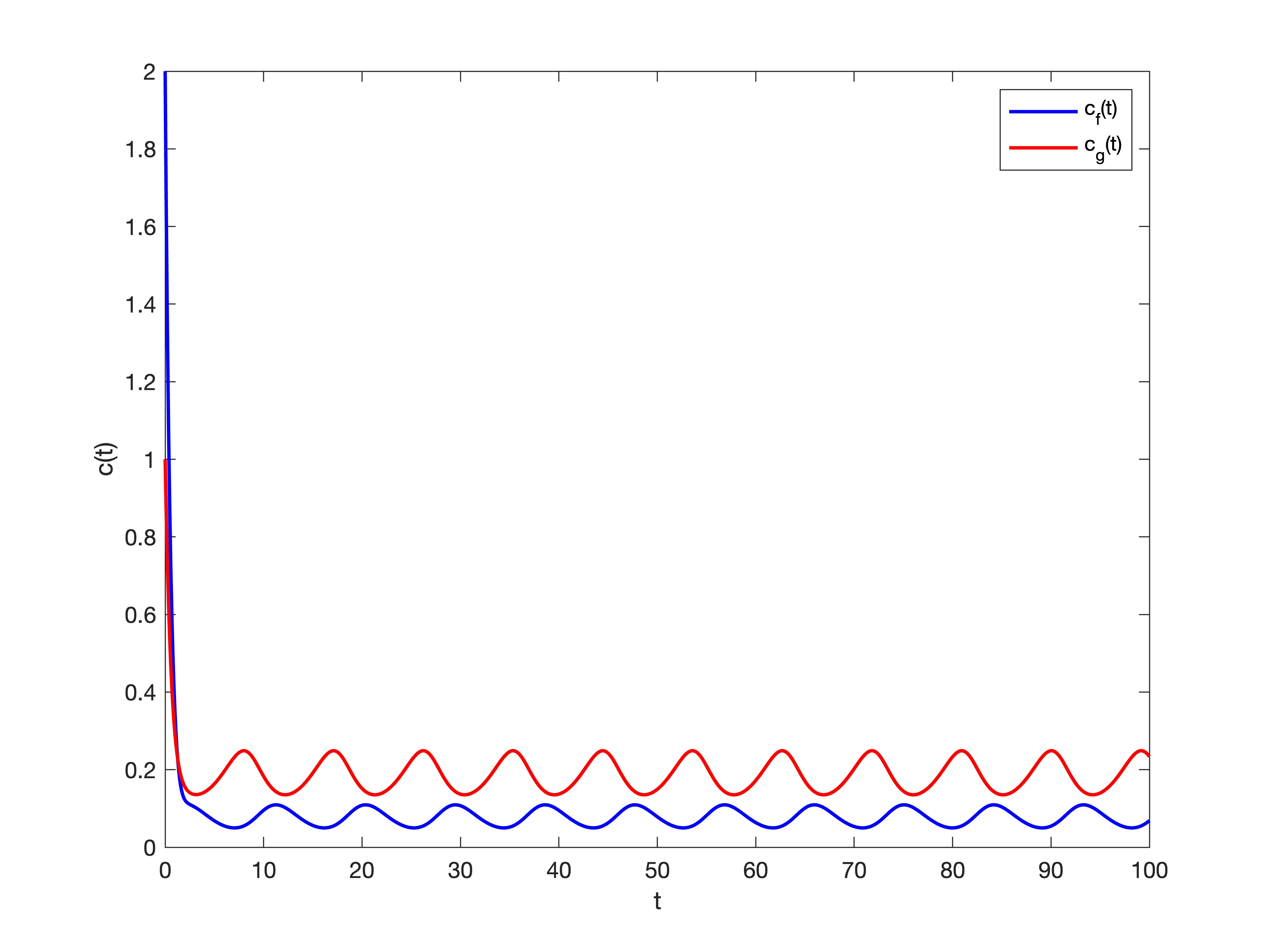}
  \includegraphics[width=0.45\textwidth]{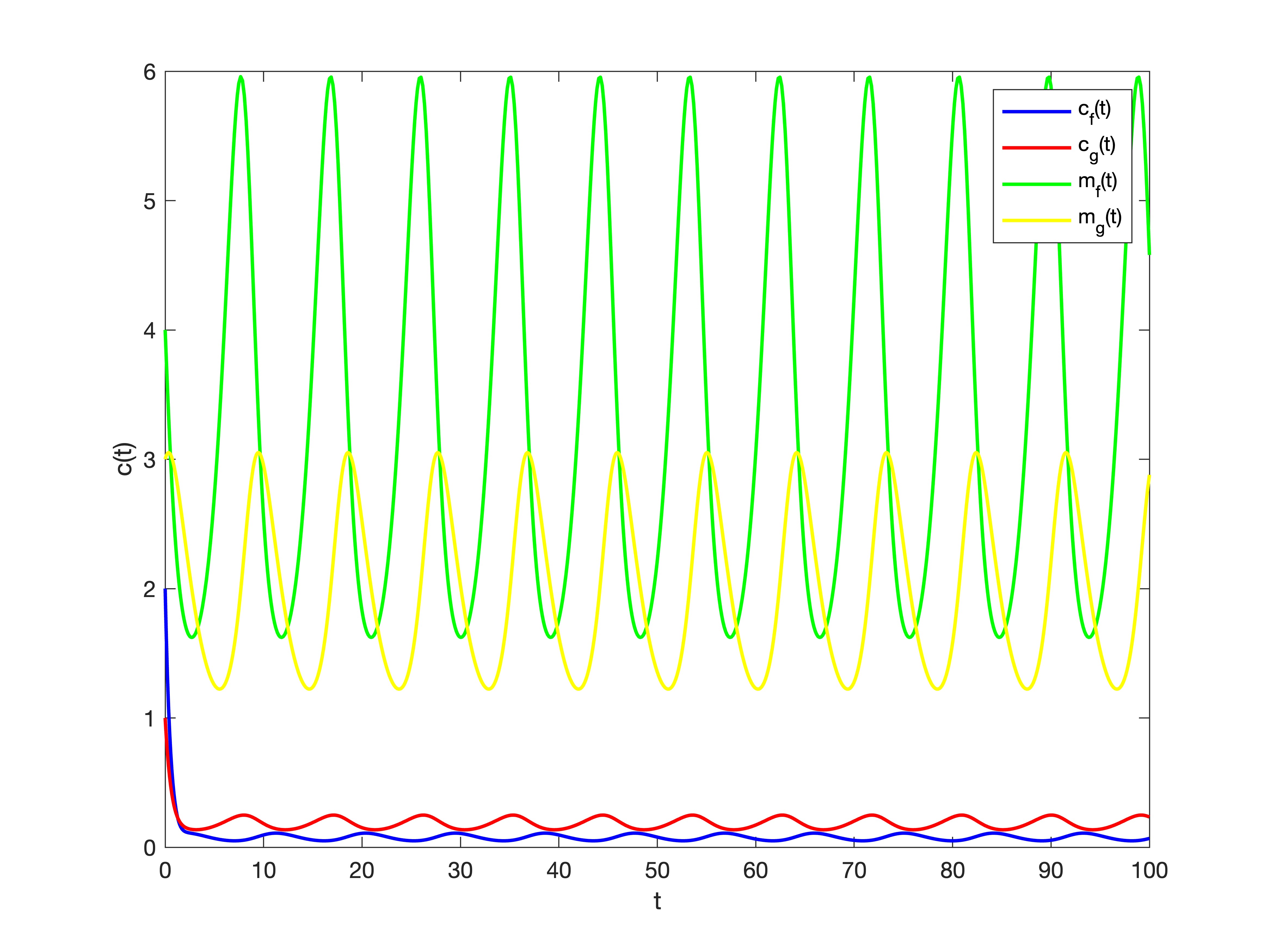}
  \caption{Evolution of the system for $p=\frac{1}{2}$ for the population and $p=1$ for the second population. Panel on the left: evolution of coefficients of variation, $c_f(t)$ and $c_g(t)$. Panel on the right: evolution of coefficients of variation, $c_f(t)$ and $c_v(t)$, along with mean values, $m_f(t)$ and $m_g(t)$. Initial data $(m_f(0), m_g(0), c_f(0), c_g(0))=(4,\, 3,\, 2,\, 1)$.}
  \label{fig2}
\end{figure}

First, we consider the case in which, in both populations, the microscopic interaction \eqref{eqmicro} is characterized by $p=\frac{1}{2}$. In this case, the quasi-equilibrium solutions are  Gamma distributions. The system of mean values is of Lotka-Volterra type, and the system of coefficients of variations is given by  \eqref{eqsysfp2}. The numerical simulations are reported in Figure \ref{fig1} and \ref{fig2}. In both cases, we can observe that the oscillations of coefficients of variation $c_f(t)$ and $c_g(t)$ occur in synchrony with those of mean values, $m_f(t)$ and $m_g(t)$. 

It is important to remark that, after a short transient, both the system of means and the system of the coefficients of variation stabilize on periodic oscillations, independently of their initial values. However, as shown in both \ref{fig1} and \ref{fig2}, through the panels on the right, 
the comparison of the amplitude of fluctuations in mean values and coefficients of variation reveals that the amplitude of the latter is greatly reduced compared to the former. The economical meaning of this reduction is clear. The system of deposit-loan volumes, while periodically subject to fluctuations, is such that the degree of inequality of the distribution of the deposit volumes, starting from an initial distribution in which they could present a high inequality, is strongly reduced. In other words, the system of deposit-loan volumes, independently of the risk factors, produces, after a short transient of time, a less unequal society. 

Further, in the case of the higher risk exponent $p=1$ for the loan volume, the inequality reduction is less marked. 

It is remarkable that lower values of the parameters $\sigma_f$ and $\sigma_g$, quantifying the degree of randomness present in the system, still determine a reduction of oscillations for both $c_f(t)$ and $c_g(t)$. 

\begin{figure}[htbp]
  \centering
  \includegraphics[width=0.45\textwidth]{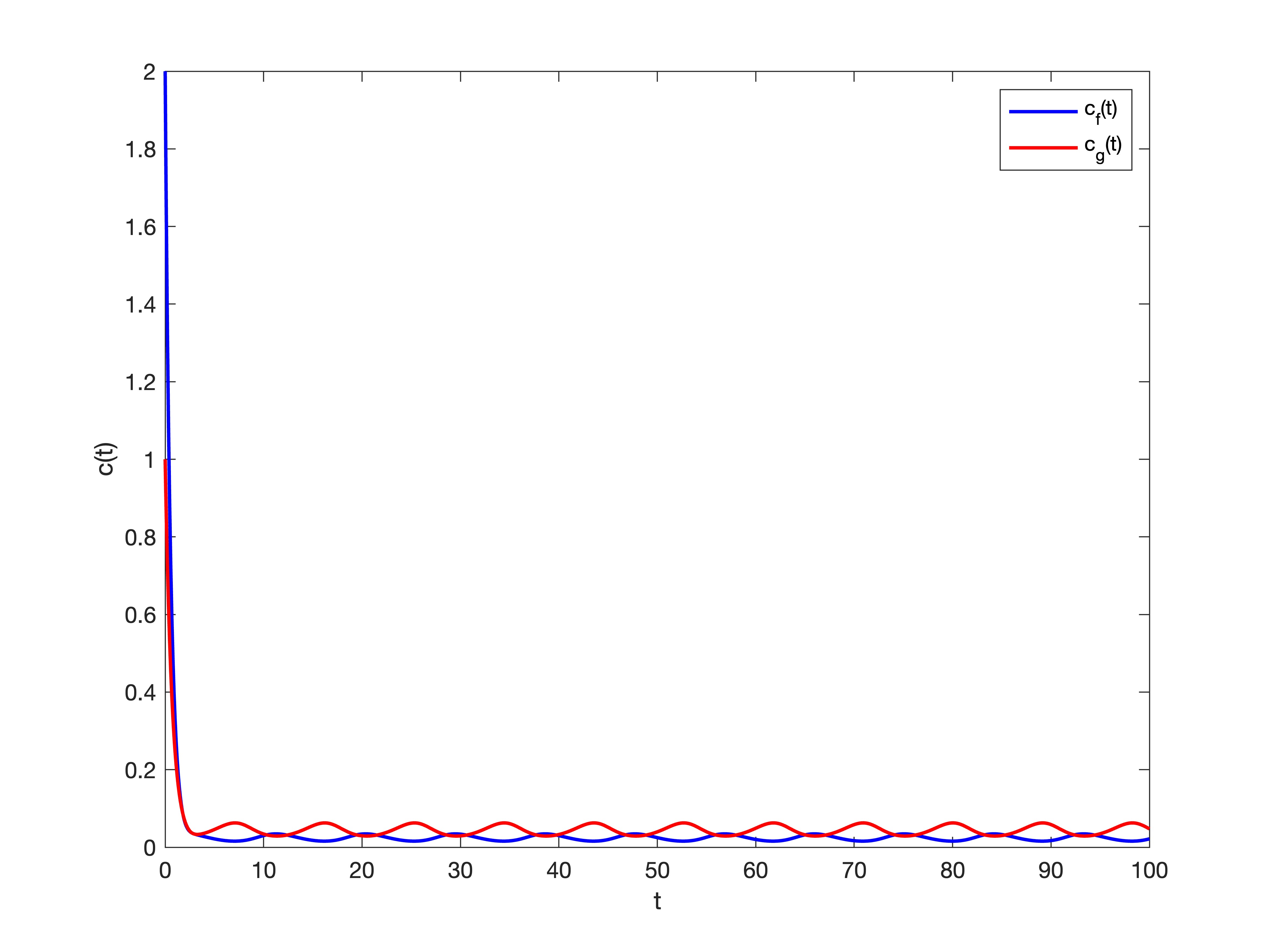}
  \includegraphics[width=0.45\textwidth]{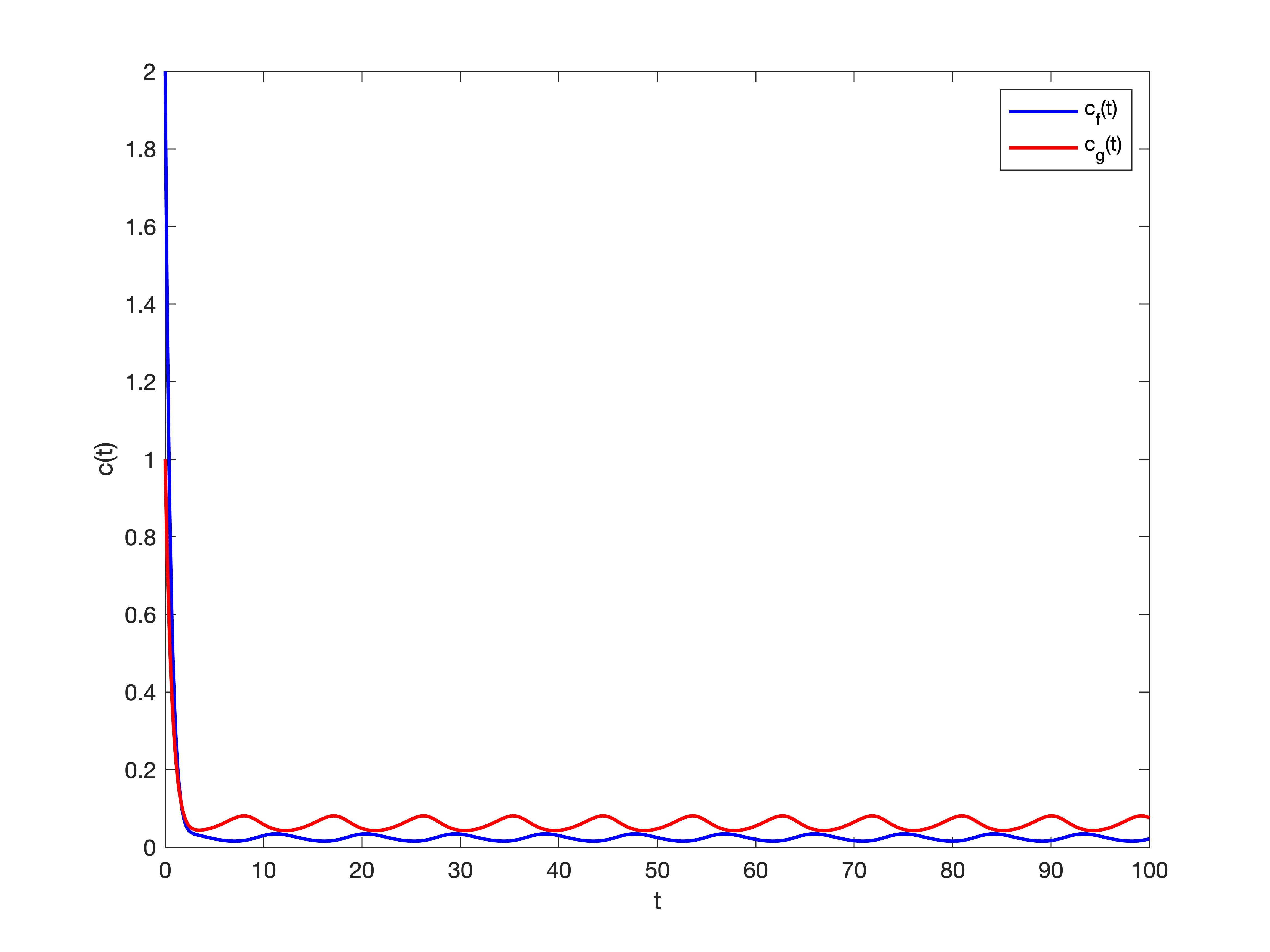}
  \caption{Evolution of coefficients of variation, $c_f(t)$ and $c_g(t)$, with a reduction of values $\sigma_f$ and $\sigma_g$ of the microscopic interactions. Panel on the right: $p=\frac{1}{2}$ for both populations. Panel on the left: $p=\frac{1}{2}$ for the first population and $p=1$ for the second population. Initial data $(m_f(0), m_g(0), c_f(0), c_g(0))=(4,\, 3,\, 2,\, 1)$}
  \label{fig3}
\end{figure}

\begin{figure}[htbp]
  \centering
  \includegraphics[width=0.65\textwidth]{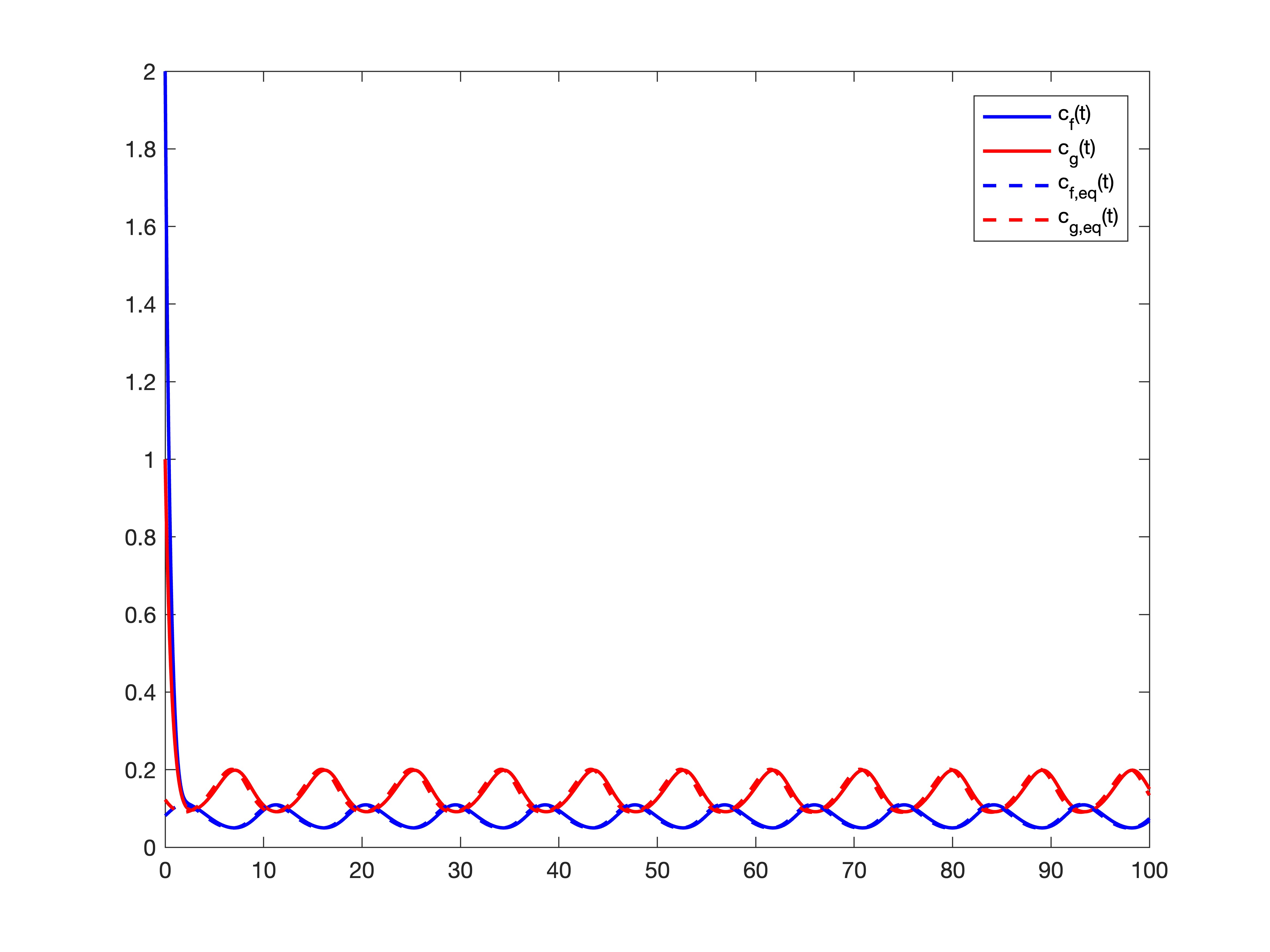}

  \caption{Evolution of the system for $p=\frac{1}{2}$ in both populations. Comparison of the coefficients of variation between the solutions of the system and the quasi-equilibrium solutions. Initial data $(m_f(0), m_g(0), c_f(0), c_g(0))=(4,\, 3,\, 2,\, 1)$.}
  \label{fig4}
\end{figure}


\section{Conclusions}

In this paper, we analyzed the evolution in time of inequality in a system of Fokker--Planck equations modeling the evolution of two populations interacting by economical reasons. In particular, a kinetic system built according to the framework proposed in \cite{SNN-2014} to model a deposit-loan interaction between a population of agents and the banks has been analyzed in terms of the coefficient of variation, a well-known measure to summarize through a scalar value the dispersion of a set of points in a statistical distribution.

The main results of this analysis are that inequality in the system tends to stabilize after a short transient of time, and that the quasi-equilibria of the Fokker--Planck system can be fruitfully used to characterize the large-time behavior only in presence of a moderate level of risk, while they can lose importance in presence of high risk. 

Numerical experiments then allow to show that this simplified model of a deposit-loan system leads to a decrease in inequality in the distribution of the volume of deposits, which can clearly be understood as a decrease in inequality in the population of agents.

The present study underlines the interest in studying the evolution of inequality in socio--economical systems characterized by the absence of a stationary state.

\vskip 3mm
{\bf Acknowledgments:}This work has been written within the activities of the GNFM
 group of INdAM (National Institute of High Mathematics). One of the authors (GT) acknowledges partial support by IMATI, Institute for Applied Mathematics and Information Technologies "E. Magenes" of CNR, Pavia.

\bibliographystyle{elsarticle-num} 
\bibliography{biblio}

\end{document}